\documentclass[twocolumn,showpacs,aps,prl,superscriptaddress]{revtex4}

\usepackage{graphicx}
\usepackage{dcolumn}
\usepackage{amsmath}
\usepackage{epsfig}

\usepackage{relsize}
\RequirePackage{xspace}
\def\babar{\mbox{\slshape B\kern-0.1em{\smaller A}\kern-0.1em
    B\kern-0.1em{\smaller A\kern-0.2em R}}}
\def\Bbar    {\kern 0.18em\overline{\kern -0.18em B}{}\xspace}
\def\Dbar    {\kern 0.18em\overline{\kern -0.18em D}{}\xspace}
\def\Kbar    {\kern 0.18em\overline{\kern -0.18em K}{}\xspace}
\def\pep2{PEP-II}
\mathchardef\Upsilon="7107
\newcommand{\optbar}[1]{\shortstack{{\tiny (\rule[.4ex]{1em}{.1mm})}
  \\ [-.7ex] $#1$}}
\def\BorBbar    {\kern 0.18em\optbar{\kern -0.18em B}{}\xspace}
\def\DorDbar    {\kern 0.18em\optbar{\kern -0.18em D}{}\xspace}
\def\KorKbar    {\kern 0.18em\optbar{\kern -0.18em K}{}\xspace}

\newcommand{\BABARPubYear}    {01}
\newcommand{\BABARPubNumber}  {23}

\newcommand{\SLACPubNumber} {9065}

\def\figurebox#1#2#3{%
    \def\arg{#3}%
    \ifx\arg\empty
    {\hfill\vbox{\hsize#2\hrule\hbox to #2{\vrule\hfill\vbox to #1{\hsize#2\vfill}\vrule}\hrule}\hfill}%
    \else
    {\hfill\epsfbox{#3}\hfill}%
    \fi}

\begin{document}

\preprint{\babar-PUB-\BABARPubYear/\BABARPubNumber} 
\preprint{SLAC-PUB-\SLACPubNumber} 

\begin{flushleft}
\babar-PUB-\BABARPubYear/\BABARPubNumber\\
SLAC-PUB-\SLACPubNumber\\ 
%hep-ex/\LANLNumber\\%[20mm]
\end{flushleft}

\title{
{\large \bf
Direct \boldmath{$CP$} Violation Searches in Charmless Hadronic 
\boldmath{$B$} Meson Decays}
}

%
%% author list as of 11-Oct-2001 (548 authors)
%
\author{B.~Aubert}
\author{D.~Boutigny}
\author{J.-M.~Gaillard}
\author{A.~Hicheur}
\author{Y.~Karyotakis}
\author{J.~P.~Lees}
\author{P.~Robbe}
\author{V.~Tisserand}
\affiliation{Laboratoire de Physique des Particules, F-74941 Annecy-le-Vieux, France }
\author{A.~Palano}
\author{A.~Pompili}
\affiliation{Universit\`a di Bari, Dipartimento di Fisica and INFN, I-70126 Bari, Italy }
% ...
\collaboration{The \babar\ Collaboration}
\noaffiliation

\date{\today}

\begin{abstract}
We search for direct $CP$ violation in charmless hadronic $B$ 
decays observed in a sample of about 22.7 million $B\Bbar$ pairs 
collected with the \babar\ detector at the PEP-II asymmetric-energy 
$e^+e^-$ collider. We measure the following charge asymmetries:
${\cal A}_{CP} ( B^{\pm}\to\eta^\prime K^{\pm} ) = -0.11\pm 0.11\pm 0.02$,\linebreak
${\cal A}_{CP} ( B^{\pm}\to\omega \pi^{\pm} ) = -0.01^{~+~0.29}_{~-~0.31}\pm 0.03$,
${\cal A}_{CP} ( B^{\pm}\to\phi K^{\pm} ) = -0.05\pm 0.20\pm 0.03$, 
${\cal A}_{CP} ( B^{\pm}\to\phi K^{*\pm} ) = -0.43^{~+~0.36}_{~-~0.30}\pm 0.06$, and
${\cal A}_{CP} ( \BorBbar^{0}\to\phi \KorKbar^{*0} ) = 
0.00\pm 0.27\pm 0.03$.
\end{abstract}

\pacs{13.25.Hw, 11.30.Er, 14.40.Nd}

%{\bf \babar\ Analysis Document $\#$310, Version 9}\\

\maketitle

The phenomenon of Charge-Parity ($CP$) symmetry violation 
has played an important
role in understanding fundamental physics since its
initial discovery in the $K$ meson system in 1964~\cite{Cronin}.
Soon after, it was recognized that the violation of $CP$ symmetry 
was one of the fundamental requirements to produce 
a matter-dominated Universe~\cite{Sakharov}.
A significant $CP$-violating asymmetry in decays of neutral 
$B$ mesons to final states containing charmonium, due to interference 
between $B^0-\Bbar^0$ mixing and direct decay amplitudes, has recently 
been observed~\cite{babarcp}. As it has now been established~\cite{directkaon}
that the $CP$-violating decays of the $K_L^0$ meson to
$\pi\pi$ final states are due to $CP$ violation in decay 
amplitudes as well as to $K^0-\Kbar^0$ mixing, 
it is topical to search for ``direct'' $CP$ asymmetries in $B$ decays,
which involve only direct decay amplitudes. These asymmetries 
are anticipated to be much larger in $B$ decays than in $K$ 
decays~\cite{Bander}.
Direct $CP$ violation would be measured as an asymmetry of $B$ 
decay rates:
%%%%%%%%%%%%%%%%%%
\begin{eqnarray}
{\cal A}_{CP}\equiv\frac{\Gamma(\Bbar\rightarrow\bar f)-\Gamma(B\rightarrow f)}
              {\Gamma(\Bbar\rightarrow\bar f)+\Gamma(B\rightarrow f)} \ .
\label{eq:acpdecay}
\end{eqnarray}
%%%%%%%%%%%%%%%%%%

Charmless $B$ meson decays are particularly interesting
processes to search for direct $CP$ violation because of the possible
involvement of penguin ($P$) and tree ($T$) amplitudes of comparable
magnitude.  Substantial $CP$ violation can thus arise in the Standard
Model through interference of these terms
\cite{Bander}:
%%%%%%%%%%%%%%%%%%
\begin{eqnarray}
{\cal A}_{CP} = \frac
{2~|P|~|T|~\sin\Delta\phi~\sin\Delta\delta}
{|P|^2 + |T|^2 + 2~|P|~|T|~\cos\Delta\phi~\cos\Delta\delta}
 \ ,
\label{eq:acpphase2}
\end{eqnarray}
%%%%%%%%%%%%%%%%%%
%%%%%%%%%%%%%%%%%%
where $\Delta\phi$ and $\Delta\delta$ are the differences
in weak and strong phases.
Because of the weak phase difference between the tree
and penguin amplitudes, ${\cal A}_{CP}$ is sensitive to 
the Cabibbo-Kobayashi-Maskawa matrix~\cite{Kobayashi}
phases $\gamma\equiv{\rm arg}\,[\,-V^{ }_{ud}V^*_{ub}\,/\,V^{ }_{cd}V^*_{cb}\,]$ 
and $\alpha\equiv {\rm arg}\,[\, - V^{ }_{td}V^*_{tb}\,/\,V^{ }_{ud}V^*_{ub}\,]$. 
The difference between the $b\to u$ tree and $b\to s$ 
($b\to d$) penguin amplitude weak phases $\Delta\phi$
is $\gamma$ ($\alpha$), as in the case of the
decays $B\rightarrow\pi K$, $\eta^\prime K$ 
($B\rightarrow\pi\pi$, $\omega\pi$).
However, large uncertainties in the strong phases,
which can be calculated by certain models,
weakens the quantitative relationship to the weak phases.
Recent calculations based on effective theory and 
factorization predict asymmetries as large as 
$\sim$${10}\%$~\cite{smphys}.

The measurement of direct $CP$ violation in pure penguin modes, 
such as $B\to\phi{K}^{(*)}$, is more sensitive to 
non-Standard-Model physics.
In the Standard Model, the lack of a tree-level contribution
results in an expected ${\cal A}_{CP}$ of no more than 
$\sim$${1}\%$~\cite{smphys}.
However, new particles in loops, such as 
charged Higgs boson or SUSY particles,
would provide additional amplitudes with different phases. Depending 
on the model parameters, ${\cal A}_{CP}$ can be $30\%$ or larger 
in such scenarios~\cite{newphys}. Complementary searches for new physics
would involve measurements of the time-dependent asymmetries in 
$\BorBbar^0$ decays to $CP$ eigenstates, such as $\eta^\prime K^0_{S(L)}$,
$\phi K^0_{S(L)}$, and $\phi \KorKbar^{*0}({\rightarrow K^0_{S}\pi^0})$.
Comparison of the value of $\sin 2\beta$ obtained from these modes 
with that from charmonium modes~\cite{babarcp}
can probe for new physics participating in penguin loops. 
In these measurements, direct $CP$ violation in the decay becomes 
highly relevant and can be studied in the self-tagging modes
discussed below.

The CLEO experiment has reported a search 
for direct $CP$ violation in $B$ meson decays to 
$\pi K$, $\eta^\prime K$, and $\omega\pi$~\cite{cleo}.
In this paper we improve the precision of the measurements
in the $\eta^\prime K$ and $\omega\pi$ modes and extend the search 
to new modes.
Measurements from \babar\ of the $B\to\pi K$ charge asymmetries 
are presented elsewhere \cite{kpipaperpl, kpipaperzr}.
Here we present measurements of the charge asymmetries 
in the following charmless $B$ decays, for which branching 
fractions have been previously reported
\cite{phipaper, etaprpaper}: 
$B^{\pm}\to\eta^\prime K^{\pm}$, 
$B^{\pm}\to\omega \pi^{\pm}$, 
$B^{\pm}\to\phi K^{\pm}$,
$B^{\pm}\to\phi K^{*\pm}$, and
$\BorBbar^{0}\to\phi \KorKbar^{*0}$.
The $B$ flavor is determined by its charge, except 
for the $\phi \KorKbar^{*0}$ final state where 
the flavor is determined from the charge of the kaon 
from the $\KorKbar^{*0}\to K^\pm\pi^\mp$ decay.

The data were collected with the \babar\ detector~\cite{babar}
at the PEP-II asymmetric-energy $e^+e^-$ collider~\cite{pep}
located at the Stanford Linear Accelerator Center.
The results presented in this paper are based on data taken
in the 1999--2000 run comprising an integrated luminosity 
of 20.7~fb$^{-1}$, corresponding to 22.7 million 
$B\Bbar$ pairs, at the $\Upsilon (4S)$ resonance
(``on-resonance'') and 2.6~fb$^{-1}$ approximately 
40~MeV below this energy (``off-resonance''). 
The $\Upsilon (4S)$ resonance occurs at the $e^+e^-$ 
center-of-mass (c.m.) energy, $\sqrt{s}$, of 10.58 GeV.

Charged particles are tracked and their momenta measured
with a combination of a silicon vertex tracker (SVT) consisting 
of five double-sided detectors and a 40-layer central drift chamber 
(DCH), both operating in a 1.5~T solenoidal magnetic field. 
With the SVT, a position resolution near the interaction point
of about 40~$\mu$m is achieved for the highest momentum 
charged particles, allowing the precise determination of 
decay vertices.
The tracking system covers 92\% of the solid angle
in the c.m. frame.
The track finding efficiency is on average (98$\pm$1)\% for momenta
above 0.2~GeV/$c$ and polar angles greater than 500~mrad.
Photons are detected by a CsI(Tl) electromagnetic calorimeter (EMC), which
provides excellent angular and energy resolution with high efficiency for 
energies above 20~MeV~\cite{babar}. The energy resolution of the EMC 
is $3\%$ and the angular resolution is 4~mrad for photons of energy 1~GeV.

The asymmetric beam configuration in the laboratory frame
provides a boost to the $\Upsilon(4S)$
increasing the momentum range of the $B$ meson decay products
up to 4.3~GeV/$c$.
Charged particle identification is provided by the average 
energy loss ($dE/dx$) in the tracking devices and
by an internally reflecting ring imaging 
Cherenkov detector (DIRC) covering the central region. 
A $K$--$\pi$ separation of better than 
four standard deviations ($\sigma$) is 
achieved for momenta below 3~GeV/$c$, decreasing to 
2.5$\sigma$ at the highest momenta in our final states. 
Electrons are identified by the tracking system 
and the EMC.

Hadronic events are selected based on track multiplicity and 
event topology. We reconstruct $B$ meson 
candidates from their charged and neutral 
decay products, including the intermediate states
$\eta^\prime\rightarrow\eta\pi^+\pi^-$ ($\eta^\prime_{\eta\pi\pi}$)
or $\rho^0\gamma$ ($\eta^\prime_{\rho\gamma}$),
$\omega\rightarrow\pi^+\pi^-\pi^0$,
$\phi\rightarrow K^+K^-$,
$K^{*\pm}\rightarrow \KorKbar^0\pi^\pm$ ($K^{*\pm}_{K^0}$)
or $K^\pm\pi^0$ ($K^{*\pm}_{K^+}$), 
$\KorKbar^{*0}\rightarrow K^\pm\pi^\mp$,
$\rho^{0}\rightarrow \pi^+\pi^-$,
$\pi^0\rightarrow \gamma\gamma$,
$\eta\rightarrow \gamma\gamma$, and
$\KorKbar^0\rightarrow K^0_S\rightarrow\pi^+\pi^-$. 
The selection requirements are identical to those used in 
the branching fraction measurements~\cite{phipaper, etaprpaper}.

Candidate charged tracks are required to originate 
from the interaction point, and to have at least 12 DCH hits 
and a minimum transverse momentum of 0.1~GeV/$c$. 
Looser criteria are applied to tracks forming $K^0_S$ candidates
to allow for displaced decay vertices.
Kaon tracks are distinguished from pion and proton tracks via a
likelihood ratio that includes $dE/dx$ information from the SVT 
and DCH, and, for momenta above 0.7~GeV/$c$, the Cherenkov angle 
and number of photons as measured by the DIRC.

We form $K^0_S$, $\phi$, $\KorKbar^{*0}$, and $\rho^0$ candidates 
from pairs of oppositely charged tracks that form a consistent vertex.
We further combine a pair of charged tracks with a consistent vertex
and a $\pi^0$ or $\eta$ candidate to select $\omega$ or 
$\eta^\prime_{\eta\pi\pi}$ candidates.
The $K^0_S$ candidates are required to satisfy 
$|m({\pi^+\pi^-}) - m_{K^0}|<$ 12~MeV/$c^2$  
with the cosine of the angle between their reconstructed flight and
momentum directions greater than 0.995
and the measured proper decay time greater than 
three times its uncertainty.

We reconstruct $\pi^0$ ($\eta$) mesons as pairs of photons,
each with a minimum energy deposition of 30~MeV (100~MeV) in the EMC.
The typical resolution of the reconstructed $\pi^0$ mass is 7~MeV/$c^2$.
A $\pm$15~MeV/$c^2$ interval centered on the nominal $\pi^0$ 
mass~\cite{pdg} is applied to select $\pi^0$ candidates. 
We combine a $\rho^0$ candidate with a photon of energy above
$200$~MeV to obtain an $\eta^\prime_{\rho\gamma}$ candidate.

We select $\phi$, $\omega$, $\eta^\prime$, and $\eta$ candidates 
with requirements on the invariant masses (in MeV/$c^2$) 
loose enough to retain sidebands for later fitting:
$990 < m(K^+K^-) < 1050$, 
$735 < m(\pi^+\pi^-\pi^0) < 830$, 
$930 < m(\eta\pi^+\pi^-) <990$, 
$900 < m(\rho\gamma) <1000$, and
$490 < m(\gamma\gamma) < 600$.
The experimental resolutions in the $K^*$ and $\rho$ 
invariant masses are negligible with respect 
to their natural widths.
The $K\pi$ invariant mass interval is $\pm 150$~MeV/$c^2$ 
for the charged and $\pm 100$~MeV/$c^2$ for the neutral 
$K^{*}$ candidates.
We require the invariant mass of $\rho$ candidates
to be between 500 and 995~MeV/$c^2$. 

The helicity angle $\theta_H$ of a $\phi$, $K^*$, or $\omega$ 
resonance is defined as the angle between the direction of one
of two daughters, or the normal to the $\omega$ decay plane,
and the parent $B$ direction in the resonance rest frame. 
To suppress combinatorial background, we require 
the cosine of the $K^{*\pm}\rightarrow K^\pm\pi^0$
helicity angle, defined with respect to the kaon,
to be greater than $-0.5$.
This effectively requires the $\pi^0$ momentum 
to be above 0.35~GeV/$c$.

We identify $B$ meson candidates kinematically
using two nearly independent 
variables \cite{babar},
the energy-substituted mass
$m_{\rm{ES}} =$ 
$[{ (s/2 + \mathbf{p}_i \cdot \mathbf{p}_B)^2 / E_i^2 - 
\mathbf{p}_B^{\,2} }]^{1/2}$ and
$\Delta E = (E_i E_B - \mathbf{p}_i 
\cdot \mathbf{p}_B - s/2)/\sqrt{s}$,
where $(E_i,\mathbf{p}_i)$ is the initial state four-momentum,
obtained from the beam momenta, and $(E_B,\mathbf{p}_B)$
is the four-momentum of the reconstructed $B$ candidate.
A quantity that is almost equivalent to $m_{\rm{ES}}$ can 
be obtained from a kinematic fit of the measured candidate
four-momentum in the $\Upsilon(4S)$ frame with its energy
constrained to that of the beam~\cite{etaprpaper}.
For signal events $\Delta E$ peaks at zero and $m_{\rm{ES}}$ 
at the $B$ mass. 
Our initial selection requires $m_{\rm{ES}}>5.2$ GeV/$c^2$
and $|\Delta E|<0.2$~GeV.

Charmless hadronic modes suffer from large background due to random 
combinations of tracks produced in the quark-antiquark  
continuum ($e^+e^-\to q\bar{q}$).  
This background is distinguished by its jet structure as 
compared to the spherical decays of the $B$ mesons produced
in the $\Upsilon(4S)$ decays. 
To reject continuum background we make use of the angle 
$\theta_T$ between the thrust axis of the $B$ candidate 
and that of the rest of the tracks and neutral clusters in
the event, calculated in the c.m. frame.  The distribution of
$\cos{\theta_T}$ is sharply peaked near 
$\pm1$ for combinations drawn from jet-like $q\bar q$ pairs, and nearly
uniform for the isotropic $B$ meson decays.
Thus we require $|\cos\theta_T| < 0.9$ (0.8 for $\phi K^{*\pm}$). 
We also construct a Fisher discriminant that combines eleven 
variables~\cite{CLEO-fisher}: the polar angles of the $B$
momentum vector and the $B$-candidate thrust axis with respect 
to the beam axis in the $\Upsilon(4S)$ frame,
and the scalar sum of the c.m. momenta of charged particles
and photon (excluding particles from the $B$ candidate)
entering nine 10$^\circ$ polar angle intervals coaxial 
around the $B$-candidate thrust axis.
Monte Carlo (MC) simulation \cite{geant} demonstrates that 
contamination from other $B$ decays is negligible.

We use an unbinned extended maximum likelihood (ML) fit to extract
signal yields and charge asymmetries simultaneously.
The extended likelihood for a sample of $N$ events is 
\begin{equation}
{\cal L} = \exp\left(-\sum_{i,k}^{} n_{ik}\right)\, \prod_{j=1}^N 
\left(\sum_{i,k}  n_{ik}\, 
{\cal P}_{ik}(\vec{x}_j;\vec{\alpha})\right) ,
\label{eq:likel}
\end{equation}
where ${\cal P}_{ik}(\vec{x}_j;\vec{\alpha})$ is the probability 
density function (PDF) for measured variables $\vec{x}_j$ of an 
event $j$ in category $i$ and flavor state $k$, and $n_{ik}$ are 
the yields extracted from the fit.
The fixed parameters $\vec{\alpha}$ describe the expected 
distributions of measured variables in each category and 
flavor state.
The PDFs are non-zero only for the correct final 
state flavor ($k = 1$ for $\Bbar\rightarrow\bar f$ and 
$k = 2$ for $B\rightarrow f$).
In the simplest case, there are two categories, 
signal and background ($i=1,2$).
The decays with the charged primary daughter 
$B^\pm\rightarrow X^0 h^\pm$ 
($h^\pm = K^\pm$ or $\pi^\pm$, and
$X^0 = \eta^\prime$, $\omega$, or $\phi$)
are fit simultaneously with two signal 
($i=1$ for $B^\pm\rightarrow X^0 K^\pm$ and 
$i=2$ for $B^\pm\rightarrow X^0 \pi^\pm$)
and two corresponding background ($i=3,4$) categories. 

We rewrite the event yields $n_{ik}$ in each category in terms of 
the asymmetry ${\cal A}_i$ and the total event yield $n_{i}$:
$n_{i1} = n_{i}\times(1 + {\cal A}_i)/2$ and
$n_{i2} = n_{i}\times(1 - {\cal A}_i)/2$.
The event yields $n_i$ and asymmetries ${\cal A}_i$
in each category are obtained by maximizing ${\cal L}$ \cite{minuit}.
The dependence of ${\cal L}$ on a fit parameter
$n_i$ or ${\cal A}_i$ is obtained with the other
fit parameters floating.
We quote statistical errors corresponding to unit
changes in the quantity
$\chi^2\equiv -2\ln{({\cal L}/{\cal L}_{\rm max})}$,
where ${\cal L}_{\rm max}$ is the maximum value
of the likelihood.
The 90\% confidence level (C.L.) limits correspond 
to a change in the $\chi^2$ of 2.69.
When more than one channel is measured for the same
primary $B$ decay, the channels are combined
by adding their $\chi^2$ distributions.

The PDF ${\cal P}_{ik}(\vec{x}_j;\vec{\alpha})$ 
for a given event $j$ is the product of PDFs 
in each of the independent fit input variables $\vec{x}_j$.
These are $\Delta E$, $m_{\rm{ES}}$, 
invariant masses of intermediate states 
($\eta^\prime$, $\omega$, $\phi$, $K^*$, and $\eta$), 
Fisher discriminant, and the $\phi$ and $\omega$ helicity angles for 
pseudoscalar-vector decays. 
For the simultaneous fit to the decays with the charged
primary daughter $h^\pm$ 
($B^\pm\rightarrow X^0 K^\pm$ and $X^0 \pi^\pm$)
we include normalized residuals 
derived from the difference between measured and expected 
DIRC Cherenkov angles for the $h^\pm$.
Additional separation between the two final states 
is provided by $\Delta E$. 
The separation depends on the momentum of the 
charged primary daughter in the laboratory and is about 
45 MeV on average varying from about 30 MeV for the highest 
momentum to about 80 MeV for the lowest momentum 
primary daughters in our final states.

For the parameterization of the PDFs for $\Delta E$, $m_{\rm{ES}}$, 
and resonance masses we employ
Gaussian and Breit-Wigner functions to describe the
signal distributions. 
For the background we use low-degree polynomials or,
in the case of $m_{\rm{ES}}$, 
an empirical phase-space function~\cite{argus}.
The background parameterizations for resonance masses
also include a resonant component to 
account for resonance production in the continuum.
In the $B$ decays to vector-vector states, the 
helicity angle distribution is the result of an 
{\it a priori\/} unknown superposition of transverse 
and longitudinal polarizations, and thus is not used
for background suppression in the fit.
For pseudoscalar-vector $B$ decay modes, angular momentum
conservation results in a $\cos^2\theta_H$ distribution 
for signal. 
The background shape is again separated into contributions 
from combinatoric background and from real mesons, both
fit by nearly constant low-degree polynomials.  
The Cherenkov angle residual PDFs are Gaussian
for both the pion and kaon distributions.
The Fisher discriminant is described by an asymmetric Gaussian 
for both signal and background. 

%%%%%%%%%%%%%%%%%%%%%%%%%%%%%%%%%%%%%%%%%%%%%%%%%%%%%%%%%
\begin{table}[h!]
\caption
{Results of the ML fits, including
number of signal events ($n_{\rm sig}$), their charge
asymmetry (${\cal A}_{CP}$), and asymmetry 90\% C.L. limits.
All results include systematic errors, which are quoted 
second after statistical errors for $n_{\rm sig}$ and 
${\cal A}_{CP}$.
}
\label{tab:results}
\begin{center}
\begin{tabular}{lccc}
\hline\hline
\vspace{-3mm}&&\\
Mode & $n_{\rm sig}$ & ${\cal A}_{CP}$ 
                               & 90\% C.L. \cr
\vspace{-3mm}&&\\
\hline
\vspace{-3mm}&&\\
$\eta^\prime K^\pm$ & 
 & $-0.11\pm 0.11\pm 0.02$ & [--0.28,+0.07]  \cr
\vspace{-3mm}&&\\
~$\eta^\prime_{\eta\pi\pi} K^\pm$ & $49.5^{~+~8.1}_{~-~7.3}\pm 1.5$ 
 & $-0.17\pm 0.15 \pm 0.01$ &  \cr
\vspace{-3mm}&&\\
~$\eta^\prime_{\rho\gamma} K^\pm$ & $87.6^{~+13.4}_{~-12.5}\pm 3.7$ 
 & $-0.05\pm 0.15\pm 0.03$ &    \cr
\vspace{-3mm}&&\\
\hline
\vspace{-3mm}&&\\
$\omega\pi^\pm$ & $27.6^{~+~8.8}_{~-~7.7}\pm 1.9$ 
 & $-0.01^{~+~0.29}_{~-~0.31}\pm 0.03$ & [--0.50,+0.46]   \cr
\vspace{-3mm}&&\\
\hline
\vspace{-3mm}&&\\
$\phi K^\pm$ & $31.4^{~+~6.7}_{~-~5.9}\pm 2.3$ 
 & $-0.05\pm 0.20\pm 0.03$ & [--0.37,+0.28]   \cr
\vspace{-3mm}&&\\
\hline
\vspace{-3mm}&&\\
$\phi K^{*\pm}$ & 
 & $-0.43^{~+~0.36}_{~-~0.30}\pm 0.06$ & [--0.88,+0.18]   \cr
\vspace{-3mm}&&\\
~$\phi K_{K^0}^{*\pm}$ & $~4.4^{~+~2.7}_{~-~2.0}\pm 0.4$ 
 & $-0.55^{~+~0.51}_{~-~0.35}\pm 0.05$ &   \cr
\vspace{-3mm}&&\\
~$\phi K_{K^+}^{*\pm}$ & $~7.1^{~+~4.3}_{~-~3.4}\pm 1.2$ 
 & $-0.31^{~+~0.54~+~0.10}_{~-~0.43~-~0.06}$ &    \cr
\vspace{-3mm}&&\\
\hline
\vspace{-3mm}&&\\
$\phi \KorKbar^{*0}$ & $20.8^{~+~5.9}_{~-~5.1}\pm 1.3$ 
 & $~~0.00\pm 0.27\pm 0.03$ & [--0.44,+0.44]  \cr
\vspace{-3mm}&&\\
\hline\hline
\end{tabular}
\end{center}
\end{table}
%%%%%%%%%%%%%%%%%%%%%%%%%%%%%%%%%%%%%%%%%%%%%%%%%%%%%%%%%

The fixed parameters $\vec{\alpha}$ describing the PDFs
are extracted from signal and background distributions from MC 
simulation, on-resonance $\Delta E$ and $m_{\rm{ES}}$ sidebands, and
off-resonance data. 
The MC resolutions in $\Delta E$ and $m_{\rm{ES}}$
are adjusted by comparisons of data and simulation 
in abundant calibration channels with similar kinematics and topology,
such as $B\rightarrow D\pi, D\rho$ with $D\rightarrow K\pi, K\pi\pi$.
The resolutions in the invariant masses of intermediate states
are obtained from inclusive particle samples.
The simulation reproduces the event-shape variable distributions
found in data.
The Cherenkov angle residual parameterizations are 
determined from samples of $\DorDbar^0\rightarrow K^\mp\pi^\pm$
originating from $D^{*\pm}$ decays.  

\begin{figure}[hbt]
\setlength{\epsfxsize}{0.9\linewidth}\leavevmode\epsfbox{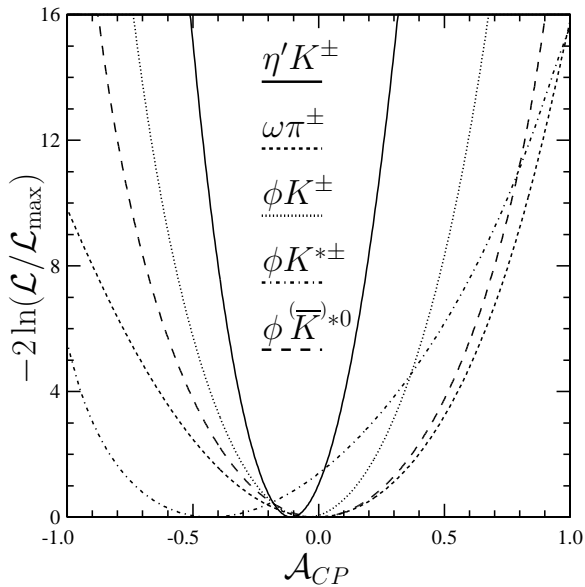}
\caption{\label{fig:acpchi} 
Dependence of $\chi^2\equiv -2\ln({\cal L}/{\cal L}_{\rm max})$ 
on ${\cal A}_{CP}$ for each of the $B$ decay modes.
}
\end{figure}

\begin{figure}[hbt]
\setlength{\epsfxsize}{0.9\linewidth}\leavevmode\epsfbox{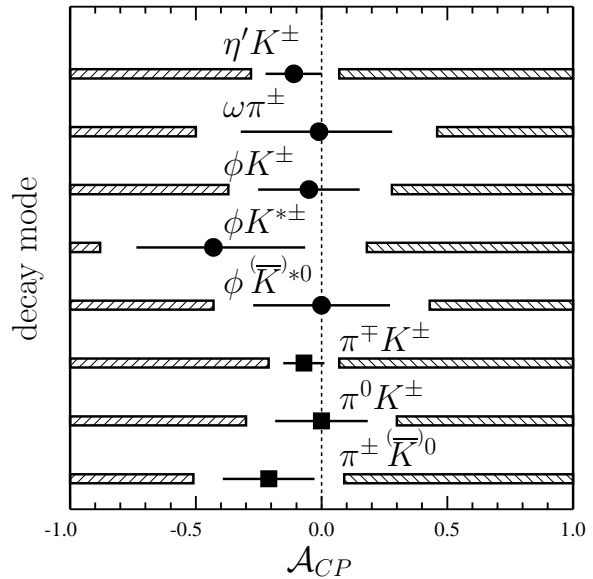}
\caption{\label{fig:visual} 
Results of ${\cal A}_{CP}$ measurements for the $B$ decay modes
presented in this paper (circles) with the $\babar$ measurements 
in $\pi K$ modes (squares) shown for comparison~\cite{kpipaperpl, kpipaperzr}.
The data sample for the $\pi^\mp K^\pm$ result is about 50\% 
larger~\cite{kpipaperzr}.
Hatched regions are excluded at 90\%~C.L.
}
\end{figure}

The results of our ML fit analyses are summarized in 
Table~\ref{tab:results}. The signal yields along with
branching fraction results have been reported earlier 
\cite{phipaper, etaprpaper}.
In all cases we find signal event yields with
significances, including systematic uncertainties,
of greater than four standard deviations,
and hence proceed with asymmetry measurements. 
The measured likelihood values are well reproduced 
with generated samples. The dependence of the $\chi^2$ on 
${\cal A}_{CP}$ for each decay mode
is shown in Fig.~\ref{fig:acpchi} and asymmetry
measurements are summarized in Fig.~\ref{fig:visual}.
We see no significant asymmetries and determine 
90\% C.L. intervals.

%%%%%%%%%%%%%%%%%%%%%%%%%%%%%%%%%%%%%%%%%%%%%%%%%%%%%%%%%
%%%%%%%%%%%%%%%%%%%%%%%%%%%%%%%%%%%%%%%%%%%%%%%%%%%%%%%%%

In the charge asymmetry measurements, systematic uncertainties
relevant to branching fraction measurements tend to cancel, 
but some level of bias is inevitable as neither the $\babar$ 
detector nor PEP-II is perfectly charge symmetric. 
However these effects are mostly very 
small for the final states considered here. 
Charge biases in track reconstruction and 
particle identification efficiency
have been studied in a sample of more than a billion charged tracks 
in multi-hadron events.  After proton and electron rejection 
we find an asymmetry in track reconstruction efficiency
consistent with zero with an uncertainty
of less than 0.01 for a wide range of momenta for tracks
originating from the interaction point. 
Taking into account particle
identification requirements similar to the ones applied to
the $K^*$ daughters, this consistency is still better
than 0.02. A $D^{*\pm}$ control sample of kaon and pion 
tracks is used to estimate systematic uncertainties 
in the asymmetries arising from possible charge biases in the
Cherenkov angle residual, which are found to be less than 0.01.

From these studies we assign a systematic uncertainty 
of 0.01 on ${\cal A}_{CP}$ for all the modes with 
a charged primary daughter: 
$B^\pm\rightarrow\eta^\prime K^\pm$, $\omega \pi^\pm$, and $\phi K^\pm$.
For the modes with a $K^*$ we account for the broader momentum spectrum 
of the charged daughters and particle identification applied to 
the kaon candidates with a 0.02 systematic error.
All measured background asymmetries in data and signal asymmetries in MC 
are consistent with zero within statistical uncertainties.

A different type of uncertainty originates in the ML fit 
from assumptions about the signal and background distributions.
In order to derive systematic errors in the event yield and its
asymmetry, we vary the PDF parameters with their respective
uncertainties.
The systematic errors in the asymmetries are found to be 
0.02 for $\eta^\prime K^\pm$ and $\phi \KorKbar^{*0}$,
0.03 for $\omega\pi^\pm$ and $\phi K^\pm$,
and 0.06 for $\phi K^{*\pm}$, the latter being dominated
by the mode with a $\pi^0$.
These systematic errors are conservatively estimated 
and can be improved with a larger data sample.

%%%%%%%%%%%%%%%%%%%%%%%%%%%%%%%%%%%%%%%%%%%%%%%%%%%%%%%%%
%%%%%%%%%%%%%%%%%%%%%%%%%%%%%%%%%%%%%%%%%%%%%%%%%%%%%%%%%

Uncorrelated (due to PDF variations) and 
correlated (due to selection requirements) 
systematic errors are treated separately in the case of multiple 
decay channels and each is convolved with the likelihood distributions
to account for all systematic effects in the result. 
The asymmetry measurement in the $\BorBbar^0\to\phi\KorKbar^{*0}$
decay mode is corrected by the inverse dilution factor $1/(1-2w)$,
where $w$, the fraction of doubly misidentified $K\pi$
combinations originating from $\KorKbar^{*0}$, is $\sim$$0.01$.
The uncertainties in the final results presented 
in Table~\ref{tab:results} are dominated by 
statistical errors.

In summary, we have searched for direct $CP$ violation in 
charmless hadronic $B$ decays observed in the $\babar$ data.
The measured charge asymmetries 
of the $B$ decays into final states $\eta^\prime K^{\pm}$, 
$\omega \pi^{\pm}$, $\phi K^{\pm}$, $\phi K^{*\pm}$, and
$\phi\KorKbar^{*0}$ are summarized in Table~\ref{tab:results}
and Fig.~\ref{fig:visual}. 
These results, along with the asymmetry measurements 
in $B\to\pi K$ modes~\cite{kpipaperpl, kpipaperzr} 
and in combination with the earlier measurements \cite{cleo},
rule out a significant part of the 
physical ${\cal A}_{CP}$ region, allowing for constraints 
on new physics models~\cite{newphys}, but are not yet of 
sufficient precision to allow precise comparison with 
Standard Model predictions~\cite{smphys}.

We are grateful for the excellent luminosity and machine conditions
provided by our \pep2\ colleagues.
The collaborating institutions wish to thank 
SLAC for its support and kind hospitality. 
This work is supported by
DOE
and NSF (USA),
NSERC (Canada),
IHEP (China),
CEA and
CNRS-IN2P3
(France),
BMBF
(Germany),
INFN (Italy),
NFR (Norway),
MIST (Russia), and
PPARC (United Kingdom). 
Individuals have received support from the Swiss NSF, 
A.~P.~Sloan Foundation, 
Research Corporation,
and Alexander von Humboldt Foundation.


\begin{thebibliography}{99}

\bibitem{Cronin}
J.~H.~Christenson, J.~Cronin, V.~Fitch, and R.~Turlay,
Phys. Rev. Lett. {\bf 13}, 138 (1964).

\bibitem{Sakharov}
A.D.~Sakharov, ZhETF Pis'ma {\bf 5}, 32 (1967);
JETP Lett. {\bf 5}, 24 (1967).

\bibitem{babarcp}
\label{ref:babarcp}
\babar\ Collaboration, B.~Aubert {\it et al.},
Phys. Rev. Lett. {\bf 87}, 091801 (2001);
BELLE Collaboration,  K.~Abe {\it et al.},
Phys. Rev. Lett. {\bf 87}, 091802 (2001).

\bibitem{directkaon}
\label{ref:directkaon}
KTeV Collaboration, A.~Alavi-Harati {\it et al.},
Phys. Rev. Lett. {\bf 83}, 22 (1999);
NA48 Collaboration, V.~Fanti {\it et al.},
Phys. Lett. B {\bf 465}, 335 (1999);
NA31 Collaboration, G.D.~Barr {\it et al.},
Phys. Lett. B {\bf 317}, 233 (1993). 

\bibitem{Bander}
\label{ref:Bander}
M.~Bander, D.~Silverman, and A.~Soni, Phys. Rev. Lett. {\bf 43}, 242 (1979).

\bibitem{Kobayashi}
\label{ref:Kobayashi}
M.~Kobayashi and T. Maskawa, Prog. Theor. Phys. {\bf 49}, 652 (1973);
N.~Cabibbo, Phys.\ Rev.\ Lett.\ {\bf 10}, 531 (1963).

\bibitem{smphys}
\label{ref:smphys}
A.~Ali, G.~Kramer, and C.-D.~L\"{u},
Phys. Rev. D {\bf 59}, 014005 (1998);
G.~Kramer, W.F.~Palmer, and H.~Simma,
Nucl. Phys. B {\bf 428}, 77 (1994).

\bibitem{newphys}
\label{ref:newphys}
I.~Hinchliffe and N.~Kersting, Phys. Rev. D {\bf 63}, 015003 (2001);
X.-G.~He, W.-S.~Hou, and K.-C.~Yang, Phys. Rev. Lett. {\bf 81}, 5738 (1998).

\bibitem{cleo}
\label{ref:cleo}
CLEO Collaboration, S.~Chen {\it et al.}, 
Phys. Rev. Lett. {\bf 85}, 525 (2000). 

\bibitem{kpipaperpl}
\label{ref:kpipaperpl}
\babar\ Collaboration, B.~Aubert {\it et al.},
Phys. Rev. Lett. {\bf 87}, 151802 (2001).

\bibitem{kpipaperzr}
\label{ref:kpipaperzr}
\babar\ Collaboration, B.~Aubert {\it et al.},
$\babar$-PUB-01/21, hep-ex/0110062, to appear in Phys. Rev. D.

\bibitem{phipaper}
\label{ref:phipaper}
\babar\ Collaboration, B.~Aubert {\it et al.},
Phys. Rev. Lett. {\bf 87}, 151801 (2001).

\bibitem{etaprpaper}
\label{ref:etaprpaper}
\babar\ Collaboration, B.~Aubert {\it et al.},
Phys. Rev. Lett. {\bf 87}, 221802 (2001).

\bibitem{babar}
\babar\ Collaboration, B.~Aubert {\it et al.},
SLAC-PUB-8569, hep-ex/0105044,
to appear in Nucl.\ Instrum.\ and Methods.

\bibitem{pep} 
PEP-II Conceptual Design Report, SLAC-R-418 (1993).

\bibitem{pdg}
Particle Data Group, D.E.~Groom {\it et al.}, 
Eur.\ Phys.\ J.\ C {\bf 15}, 1 (2000).

\bibitem{CLEO-fisher}
CLEO Collaboration,
D.M.~Asner {\it et al.}, 
Phys.\ Rev.\ D {\bf 53}, 1039 (1996).

\bibitem{geant}
The \babar\ detector Monte Carlo 
simulation is based on GEANT:
R.~Brun {\it et al.}, CERN DD/EE/84-1.

\bibitem{minuit}
F.~James,
CERN Program Library, D506.

\bibitem{argus}
ARGUS Collaboration, H.~Albrecht {\it et al.}, Phys.\ Lett.\ B {\bf 241}, 278 (1990).

\end{thebibliography}
\end{document}